# Thermoconvective structure dynamics in horizontal Homeotropic Nematics films heated from above

I. Bove[1] & J. Salán[2]

[1]Instituto de Física, Facultad de Ingeniería, University of the Republic, Uruguay

[2]Dpt. E.C.M., Fac. Physics, University of Barcelona, Spain

E-mail: italo@fing.edu.uy

**Abstract**

Homeotropic Nematic Liquid Crystal heated from above present convective Rayleigh-Benard instability for applied thermal gradients greater than $\Delta T_c$. This threshold increase with the intensity of applied external magnetic field parallel to the initial molecular orientation (vertical). We studied the different patterns which exist near to and away from the threshold, such as hexagons and squares, and the transitions between them. The study of patterns evolution away the threshold shows coexistence between patterns and a global system dynamics. Also we studied how this general dynamics change with the inclination of the cell, where for very small angles (1 degree) the hexagons and squares are transformed into rolls.

## 1. Introduction

It is well-established that when a horizontal film of Homeotropic Nematic Liquid Crystal is heated from above, a convective instability appears for moderate thermal gradients applied in the gravitational direction [1]. The underlying mechanism of the linear instability modes can be described as the following: the fluctuations of initial molecular orientation (defined by a director field **n(r)** with $|\mathbf{n}(\mathbf{r})|=1$) induces a thermal fluctuation via a heat focusing-defocusing effect. This is because of the anisotropy of the thermal diffusivity $\kappa$ of the Nematic phase ($\kappa_{\parallel}/\kappa_{\perp} \approx 1.3$ where $\parallel$ and $\perp$ refer to directions of heat flux with respect to **n**) [2]. Buoyancy forces and convective flow occur under the influence of the density modification. Also, distortion of the director is induced by shear gradients reinforcing the initial director fluctuation. The characteristics of the linear convective Rayleigh-Benard instability are strongly modified by including the director fluctuation because of its long diffusive relaxation time compared with those of heat and vorticity. Linear and weak non-linear models have been developed in order to explain this instability [3] [4] [5] [6].

The application of the external magnetic field H, parallel to the initial molecular Homeotropic orientation (in the vertical direction), reinforces the elastic effects (as shown elsewhere [7]). Then, the threshold increases with the intensity of applied H. A diagram showing the variation of threshold with the magnetic field was made in Salan & Guyon, mapping the regions where different spatial patterns appear. So, firstly, for null or small H, the cross-roll patterns appear near the threshold. Secondly, for higher H, the hexagonal patterns appear near the threshold as a consequence of non-Boussinesq effects, such as the strong variations of kinematic viscosity [8].

Until recently, studies on pattern formation and turbulence driven by thermal gradients have been typically focused on Rayleigh-Bénard convection (for a review of it see [9]) with or without including the surface tension (Marangoni effect). Concerning the latter, for isotropic fluids, when surface tension dominates buoyancy, both theoretical and experimental studies have been published. This is not the case of homeotropic Nematics, where a few number of theoretical [3] [10] [11] and experimental [6] [7] works have been developed. In general, the experimental effort was restricted to large aspect ratios in hexagonal convection in the vicinity of thresholds [12] [13]. The non-linear studies in Rayleigh-Benard-Marangoni convection can be grouped into bifurcation analysis based on amplitude equations [14] [15] [16] [17] and direct numerical simulations of the balance equations [18] [19]. These works predict the same hierarchy of transitions when the thermal gradient is increased, starting with hexagons at onset and going to square or roll patterns for larger gradients. Some experiments [18] show that square patterns are solutions away from the threshold, and also, theoretical models have been developed predicting these solutions. The competition between hexagons and squares has been theoretically analysed and experimentally studied [19]. In Homeotropic Nematics, our previous study [6] examined the hexagon-square transition experimentally.

Here we report experiments on Rayleigh-Benard convection in Nematic Liquid Crystals heated from above. Firstly, we reproduce previous experimental results for different thickness of Nematic films with applied external H parallel to the director **n**, verifying the existence of regions where hexagons and squares appear. Secondly, we follow our previous experiments [7] studying the evolution of the different pattern domains for large time interval. We show the existence of global system dynamics measuring the average pattern velocities which change with the external parameters, such as the thickness of the film, H, and thermal gradients. A first esteem of the relative densities of squares in front of hexagons is made for several values of parameters H and applied ΔT. Also, we examined the change in the dynamics when the cell is tilted, arriving to the transformation of the existing structures into rolls. Finally, we discussed our results comparing them with theoretical predictions (cited above) and previous experiments in three kinds of fluids: isotropic, liquid crystals, and ferrofluids.

## 2. Experimental Set-up

Figure 1 shows the Experimental Set-up, viewed laterally. The liquid crystal used is MBBA, which presents the Nematic phase between 18º and 42ºC. It is placed between two parallel rectangular glass plates (dimensions $d_x$=9cm and $d_y$=3cm). The thermal conductivity of the plates is six times as much the one of the liquid (see for instance, [7]). The cell is closed sideways by an insulating Teflon rectangle, which also serves as a spacer. The thickness of the Nematic film was $d_z$=0.8mm and 1mm. So the aspect ratio was 90 or more.

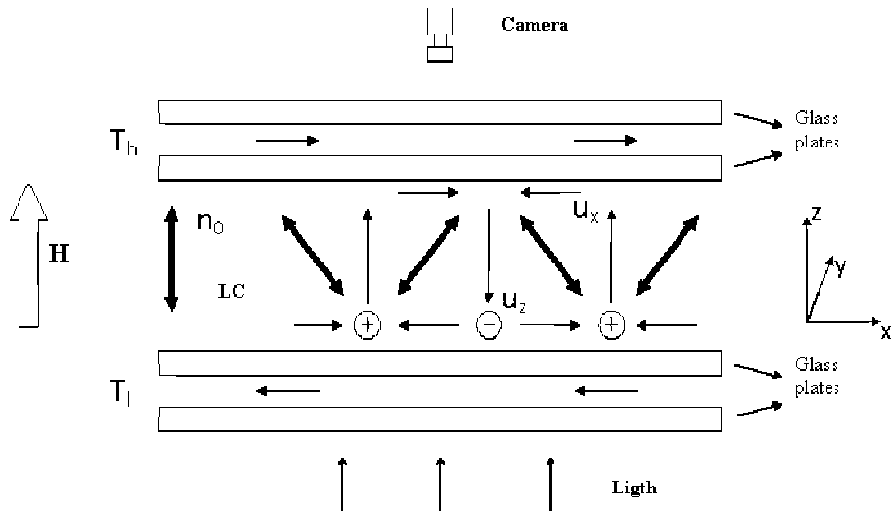

**Fig. 1.** Schematic cross-section of the experimental cell. A thick layer or homeotropic MBBA is sandwiched between two transparent glasses plates; the thickness of the cell is defined by a Teflon spacer. The double arrows represented the fluctuation of the director and the temperature fluctuation induced by 'heat focusing' is given by ⊕ and ⊖.

Two independent water circulations create the temperature gradient over the outer faces of the cell connected to thermal regulated baths. The baths have a refrigeration circuit with oscillations smaller than 0.01ºC. The temperature difference ΔT is kept to ±0.05ºC over 24h. ΔT is read by a system of differential thermocouples (wire diameters 0.12 mm) placed in outer and inner glass plate faces. Several thermal gradients (vertical and horizontal) were controlled and registered by a scanning system connected to a computer. The horizontal thermal gradients were smaller than 0.2ºC over the largest length $l_x$. The temperature drops are estimated to be less than 2% of ΔT. The control of the horizontal thermal gradients was important in order to study the dynamics of spatial patterns, as we discuss later.

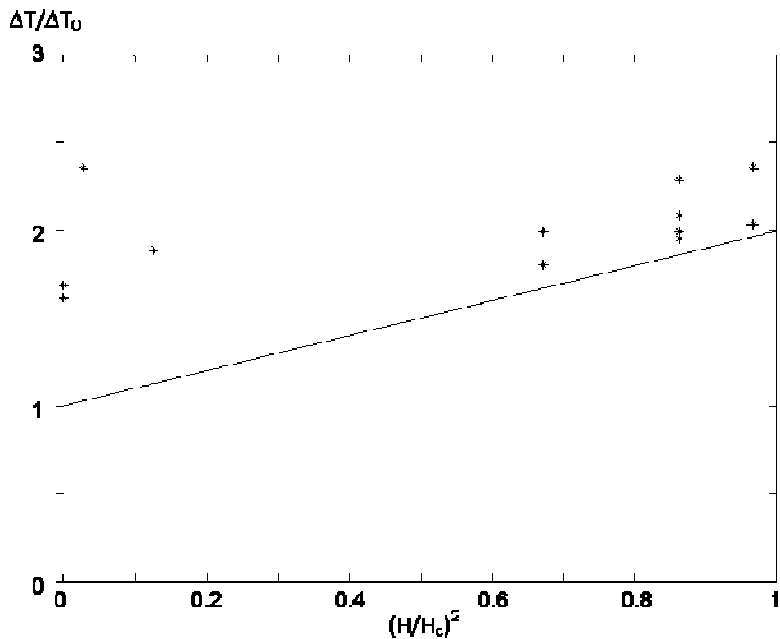

**Fig. 2.** Area of studies respect H and ΔT. $\Delta T_0$ correspond to the critical difference of temperature for H=0 and $H_c$ to the critical Freederickz. Starts (*) correspond to the points (away of the threshold) where evolution of patterns was observed. Comparation is made with theoretical linear law for the threshold (solid line).

The Homeotropic alignment of the Nematic is obtained by pre-coating the inner face of the glass plates with a thin lecithin film. A system of Helmholtz coils (range: 0-800 Gauss, with a precision better than 5 G) is used in order to apply an external magnetic field H in vertical direction.

Detection of convection makes use of the large birefringence that accompanies the periodic distortion of molecular orientation above the threshold $\Delta T_c$. The dark lines and points correspond to the regions where the fluid moves vertically. The patterns are visualised directly increasing the image contrast by polarizers. A CCD camera and a Matrox system of process and analysis is used to register time controlled sequences of images for different applied $\Delta T$ and H. In all the experiments, the applied temperature differences (close to or far from the threshold) were made by increasing $\Delta T$ by 0.2ºC steps. The time between steps was 30min. When it reached the selected convective regime (for $\Delta T$, H fixed), times greater than 12 h were attained before we studied the evolution of spatial patterns.

## 3. Results

In order to better describe the experimental results, we present it in 3 parts.

### 3.1 Thresholds and Patterns

Although we did not systematically reproduce previous experimental and theoretical research [1] [7] [20], we verified several key points. In Figure 2 we plot the variation of threshold with H in supercritical states compared with the smooth non-linear theory [10]. The solid line in Figure 2 corresponds to the theoretical predictions. As expected, for small $\Delta T_c(H)$ near the threshold, the convective structures are rolls evolving to cross-roll patterns similarly as were described in previous works [6] [7] [21]. For greater $\Delta T_c$ (increasing H), hexagonal patterns appear when the threshold reach the region where non-Boussinesq effects are important. The viscosity coefficients suffer variations greater than 50% for the applied gradients, as reported and discussed elsewhere [7] [8]. Also in the region of small $\Delta T_c$ and H, when $\Delta T$ is increased slowly (H constant) the cross-roll pattern is reinforced because the non-linear effects produce stronger distortions of **n** (the visual image is a square pattern). Finally, for large thermal gradients, the squares evolve to hexagons.

Our experiments focus on the appearance, variations, and evolutions of hexagonal and square patterns. In order to determine the states, we define $\varepsilon=(R-R_c)/R_c$ (where R and $R_c$ are the Rayleigh and critical Rayleigh number respectively which depend on the intensity of applied H), and $h=(H/H_c)^2$ (here, $H_c$ is the critical Freedericksz magnetic field [6]). In figure 3 an area of uniform hexagons can be observed, which is found in a region with large H and small distance to threshold. In Figure 4 there is a square region with many defects, since it was not possible to find homogeneous square structures. Both figures also shows the Fourier Transform of the structures. This Transform allows checking the homogeneity of the structures by detecting variations in the direction of the wave vector λ. To compare these structures the distances between the centers of the squares ($d_S$) and hexagons ($d_H$) were measured. These are expressed in non dimensional form dividing them by the thickness of the cell, which is the natural length of the experiment. From these, we calculated wavelengths λ: for the squares, $\lambda_S$ is simply $d_S$, but for the hexagons is somewhat different (see [22]):

$$d_S = 1{,}80 \pm 0{,}05; \quad \lambda_S = d_S; \quad A_S = d_S^2 = 3{,}24;$$

$$d_H = 2{,}15 \pm 0{,}05; \quad \lambda_H = 1{,}86; \quad A_H = (\sqrt{3}/2)\, d_H^2 = 4{,}00.$$

The equality of the wavelengths of both structures draws our attention, within the range of error in the measurements. These measures were corroborated with the study of the Fourier Transforms of the different regions of the cell that had an area with homogeneous structures.

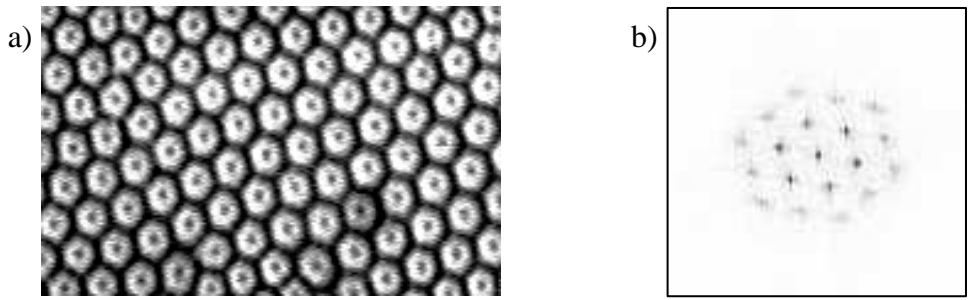

**Fig. 3.** a) An example of regular hexagons for h=1,59 y ε=0,04. b) Its corresponding Fourier Transform.

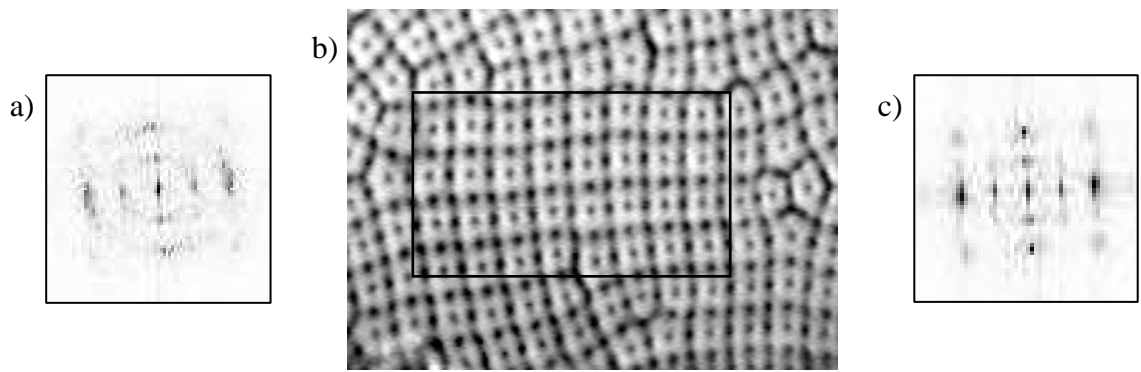

**Fig. 4.** a) Fourier Transform of the entire region of the image (b). b) Region of squares for H=0 and ε=0.6. c) Fourier Transform of the inner frame of the image (b)

Before we could begin to study the diagram of different patterns in supercritical states, we developed experiments fixing two kind of initial points in a diagram (ΔT, H): convective points away from the linear threshold (for small H) and convective points near the onset (for strong H). We first considered the case of strong H (where non-Boussinesq effects are important). Near the threshold, a system of clusters with regular hexagon patterns is obtained. When ΔT is increased (0.2 ºC each 2-3 hours), domains of squares appear around the point or line defects of the initial pattern as well as on the borders. The frontiers of the hexagon or square domains are formed by pentagon files (also named grain boundaries). Though the orientation of the three modes of hexagon patterns can change for different clusters, typically this orientation is correlated to the frontiers of the clusters, which is defined as Type I fronts [14]. In our experiments, it was not possible to obtain regular patterns probably as a consequence of the geometry of the box (rectangular) and aspect ratios (greater than 90) used. These kinds of effects had been discussed in other works ( [13] [23] and refs. there in). Also, the dynamical effects which are present in our experiments for supercritical states (described and discussed later) make it impossible to obtain regular patterns over the cell. Figure 4 shows pairs of penta-heptagons, and grain boundaries can be observed. These are also shown in electrodynamical convection in Liquid Crystals [24] [25] or ferrofluid instabilities [26] [27]. When ΔT is increased, a significant amount of point defects and grain boundaries is observed and produces an increase of square domains and other domains as well. For time delays over 100 hours, we could not obtain regular square patterns over the whole cell, most likely as a consequence of the geometrical conditions and the dynamics of the system, as stated before. The hexagon-square transitions and the creation of squares are directly related to the dynamical effects and described in paragraph 3.2.

In the case when H is smaller, we also start near the onset where the pattern was convective rolls evolving to cross-rolls or squares. In previous experiments [1] [7] the appearance of squares was reported at onset. In our experiments, when the first instability is forming hexagons the squares pattern appears away from the threshold. When the first instability is forming squares and if ΔT is increased (with constant H), the hexagon domains appear (similar situation was analized theoretically by Madruga et al. [19]), which are a kind of deformation of the previous square pattern formed on the lateral boundaries. Also, in this case, the separation fronts between patterns was the Type I fronts.

We observed, over different values of ΔT (>ΔT$_c$) and large time interval, a coexistence between two kinds of patterns. Patterns evolved continuously with the dynamics changing the ratio of squares/hexagons S with H and ΔT. The ratio S and its evolution was analysed for several values of H and applied thermal gradients. In all cases considered the ratio S changes for several days (4-6 days) before it reaches a final stationary state. Three experimental situations were considered for observation times longer than 10 days. For ε=0.13 and h=0.82, S was 0.2. For a small increase in applied H (h~0.95) and ε=0.25, S~0.07. Also, for h=0 and ε~0.65 the ratio S was 0.42. This result shows that in the region where the first instability is squares, for an applied null magnetic field, it is necessary to strongly increase the thermal gradient in order to obtain a relative density of hexagons occupying almost one half of the cell. For small magnetic fields, the density of hexagons grew more quickly, although a quantitative result was not possible because the observation times did not allow us to assure a stationary state. We believe, from these results and other qualitative evidences in our experiments, that the influence of H over the existence and amount of hexagon clusters is more important than those of increased thermal gradients (or ε).

**3.2 Dynamics of the structures**

For several considered convective states consisting of hexagons and squares with fixed ΔT and H near to or away from the threshold (represented in figure 2), we have detected slow displacements of convective cells and domains over the entire box. Differential thermocouples were placed on the cell borders to control possible horizontal temperature gradients during the time intervals of image acquisitions (100 h or more). Thus, any relevant horizontal ΔT was measured. In Figures 5 and 6 a series of photographs of the cell with and without magnetic field respectively can be observed. There you can clearly see its overall dynamics. In the bottom of both figures the arrows represent the average speeds of the structures calculated for different points in the cell, and the inner frame represents the area that is observed in each of the photograms. The velocities were calculated for the same time intervals separating the images from each other, so the arrows represent the movement of the structures photographs to photographs.

**Table I.** Typical velocities and their error, for the most relevant cases, in function of the parameters of the problem.

|    | Thickness (mm) | ΔT (°C) | $(H/H_c)^2$ | ε | mean v. (mm/day) | error mean v. | v. max. (mm/day) | error v. max. |
|----|----|----|----|----|----|----|----|----|
| **a)** | 0,8 | 17,4 | 0,00 | 0,66 | 2,99 | 0,56 | 5,32 | 0,69 |
| **b)** | 1,0 | 8,3  | 0,00 | 0,52 | 1,70 | 0,19 | 2,56 | 0,15 |
| **c)** | 1,0 | 12,1 | 0,03 | 1,16 | 5,77 | 0,82 | 8,32 | 0,75 |
| **d)** | 1,0 | 8,9  | 0,74 | 0,06 | 0,99 | 0,36 | 3,26 | 0,82 |
| **e)** | 1,0 | 9,9  | 0,79 | 0,15 | 1,61 | 0,23 | 2,77 | 0,18 |
| **f)** | 1,0 | 13,0 | 1,13 | 0,3  | 4,88 | 0,33 | 6,33 | 0,32 |

Until now, there has not been a systematic study of the pattern dynamics. However, we can provide some results on this topic. Measurements of the displacements were made for the variables of the problem. Table I shows the typical velocities and their error, in relation to the applied ΔT and H. Increasing the intensity ΔT, for the same H, the displacement velocities of patterns increase too. Also, for a constant ΔT, increasing H, the velocities decrease. These behaviors are consistent with the results of a perturbative analysis where the fact that amplitudes grow with distance to convective threshold Tc is obtained [13], which in our case depends on both T and H. It is also observed that, for similar relative distances ε to convective threshold, the higher values H applied, the greater the speed of the structures. This relationship is more difficult to analyze because the influence of H on convective dynamics for nonlinear behavior has not been studied so far. Clearly H plays a role in convection, and not only ΔT, and thus the assumption that the CL is a non-Boussinesq fluid should be supplemented with the fact that the problem depends on the magnetic field.

Also, there are other results. The velocities change clearly with the thickness of the Nematic film, being higher for smaller thicknesses (for all the velocities diagrams, see [28]). In all the cases it can be seen that the trajectories are not rectilinear, describing curves with a large radius, and these radius are smaller for smaller thicknesses. The domains of hexagons and squares travel with the same velocities as individual cells and so, the separation fronts move to the same velocity as the clusters.

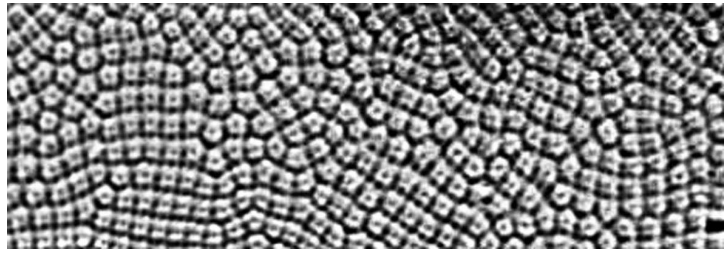

T=00 h.

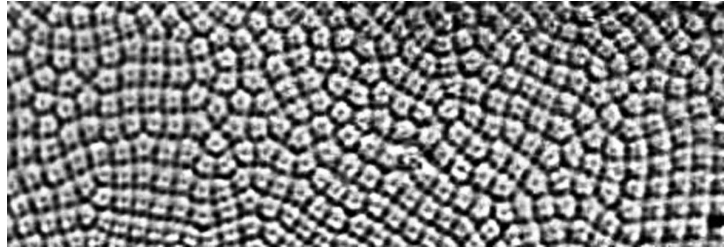

T=24 h.

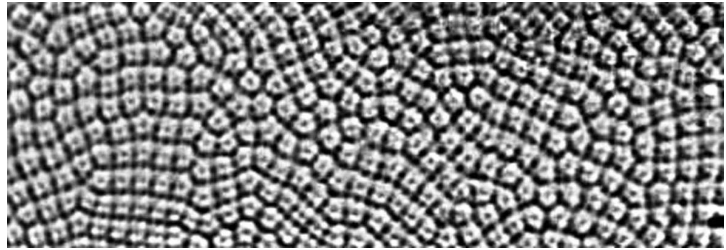

T=48 h.

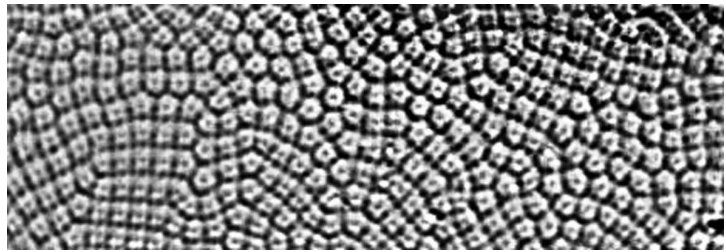

T=72 h.

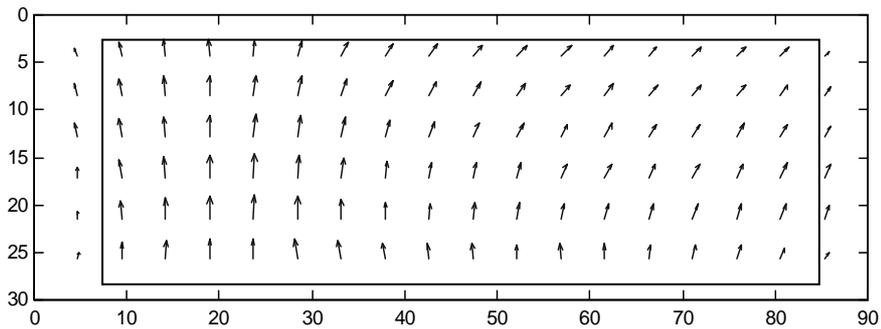

**Fig. 5.** A sequence of pictures each 24 hours in the experiments conditions: H=0 G, ε=0,62 and d=1 mm. The variation of the structure positions in the different regions of the cell can be observed.

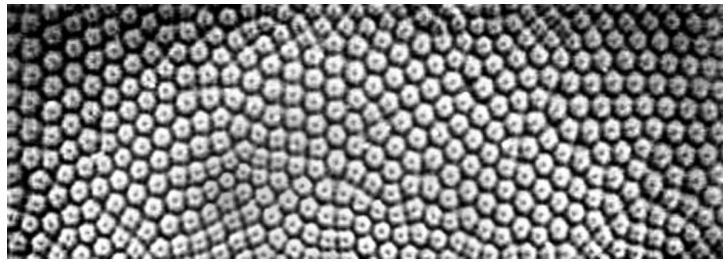

T=00 h.

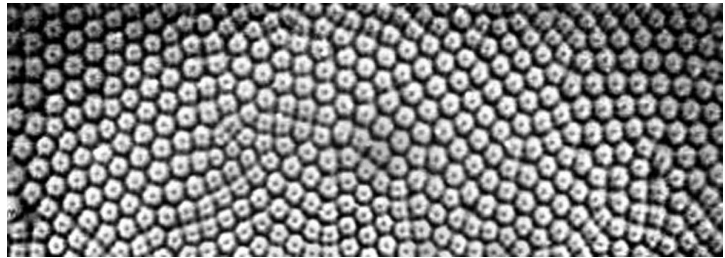

T=12 h.

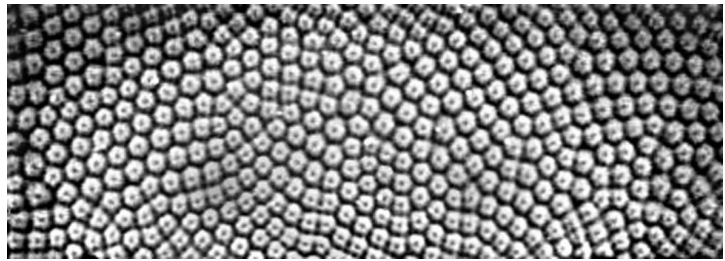

T=24 h.

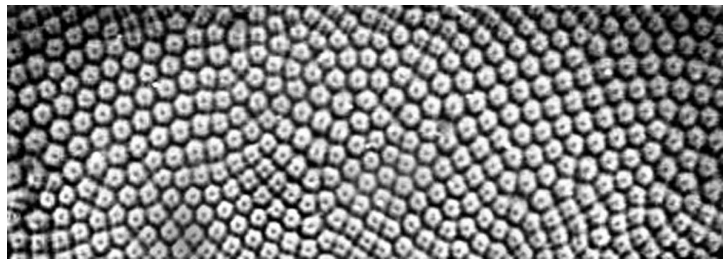

T=36 h.

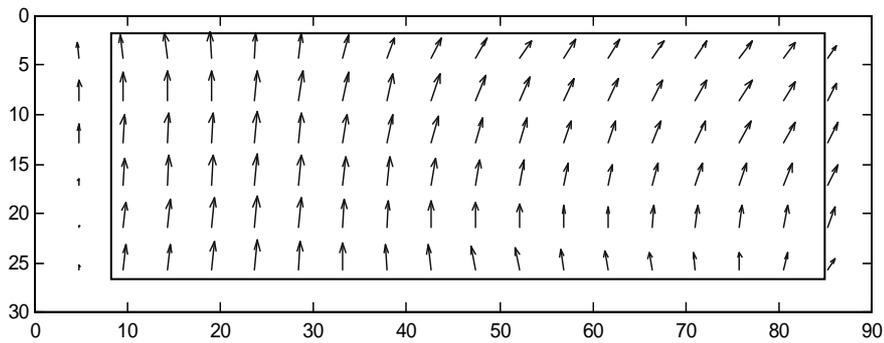

**Fig. 6.** A sequence of pictures each 12 hours for the experiments conditions: H=79 G, ε=0,27 and d=1 mm. We can see that the structure positions variation is slower than the one in fig. 5. Also, since ε is smaller, there is more area of hexagons.

However, some singular points could be observed where the structures flow around these fixed points (or with small velocities compared to the dynamics of structures) during long time periods. All the cases belonged to penta-heptagon defects. Near the coins of the rectangular box, more complicated movements exist describing trajectories almost circular or spiral, similar to the vortices in fluid flows past obstacles, although larger time periods are necessary to complete the trajectories in this case.

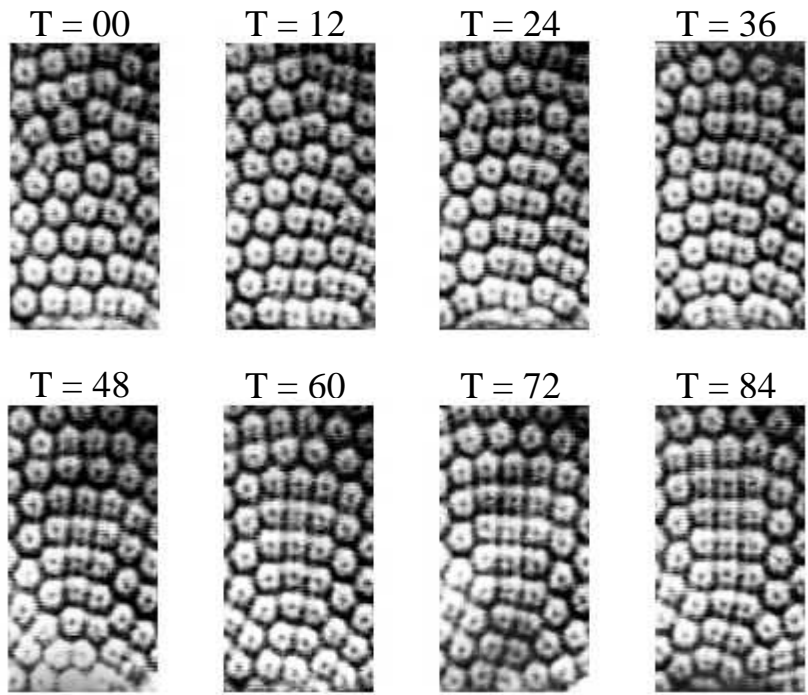

**Fig. 7.** Evolution of a cluster from hexagons to squares with an intermediate state of pentagons line. The sequence of pictures were taken each 12 hours in the experiments conditions: H=66 G, ε=0,13 and d=1 mm.

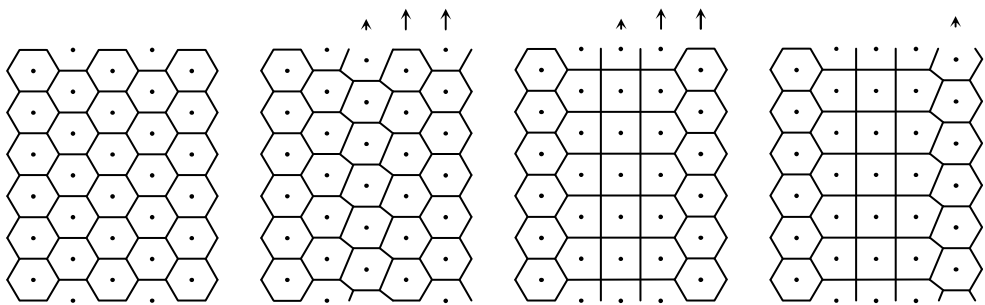

**Fig. 8.** A schematic geometrical design for the evolution of a cluster from hexagons to squares produced by the difference in the velocities of adjacent rows. The frontier between domains is type I, as it can be seen by the resulting pentagon line.

In the region where the first instability is forming hexagons, if we placed the relevant parameters (ΔT, H) where squares appear, the system dynamics follow trajectories as it is shown in Figure 6. We observed that in the general dynamics of the structures, hexagon-square transition occur continuously forming new clusters from the previous pattern. In fact, in our experiments it was impossible to obtain a regular pattern of hexagons or squares. Both kinds of patterns coexist forming domains separated by grain boundaries. The number of domains changes with the parameters of the system ΔT, H, and thickness, as described above.

A typical mechanism for these hexagon-square transitions is illustrated in Figure 7. It shows the creation of a cluster of squares from a hexagon domain with a defect. The traslational motion of hexagon clusters in different directions around a defect produce a sliding of a row of hexagons respect to the adjacent cluster, generating one row of rectangles which evolves just to form a final stable row of squares. Figure 8 shows this hexagon-square transition schematically. More generally, small divergences in the trajectories of adjacent clusters imply the disappearance of one of three modes corresponding to a hexagonal pattern and so, rows of rhombic or rectangular structures appear, which evolve to squares pattern.

Another mechanism which exists in our experiments is that when the parameters of the problem are changed, the relative density of hexagons - squares S varies via the creation of squares (or hexagons) in one border of the cell, while on the opposite border hexagons (or squares) are destroyed. This mechanism is associated with the overall dynamic, described above, which moves the structures along the entire cell, causing structures at the edges constantly being created and destroyed.

### 3.3 Convection in inclined planes

To try to understand better the general dynamics in our LC cell, we examined the change in the dynamics when it is tilted, which is a study that has never been done before. We found that even for very small angles (0.25 degrees or less) there are significant changes in the dynamics, and for angles of 1 degree only, the hexagons are transformed into rolls. Table II shows the various cases studied, showing the inclination on each one, and the speeds of the structures obtained, both average and maximum (with their errors). To refer in an abbreviated form to the inclinations, $I_r$ is defined as the angle of rotation about an axis r perpendicular to the plane of rotation.

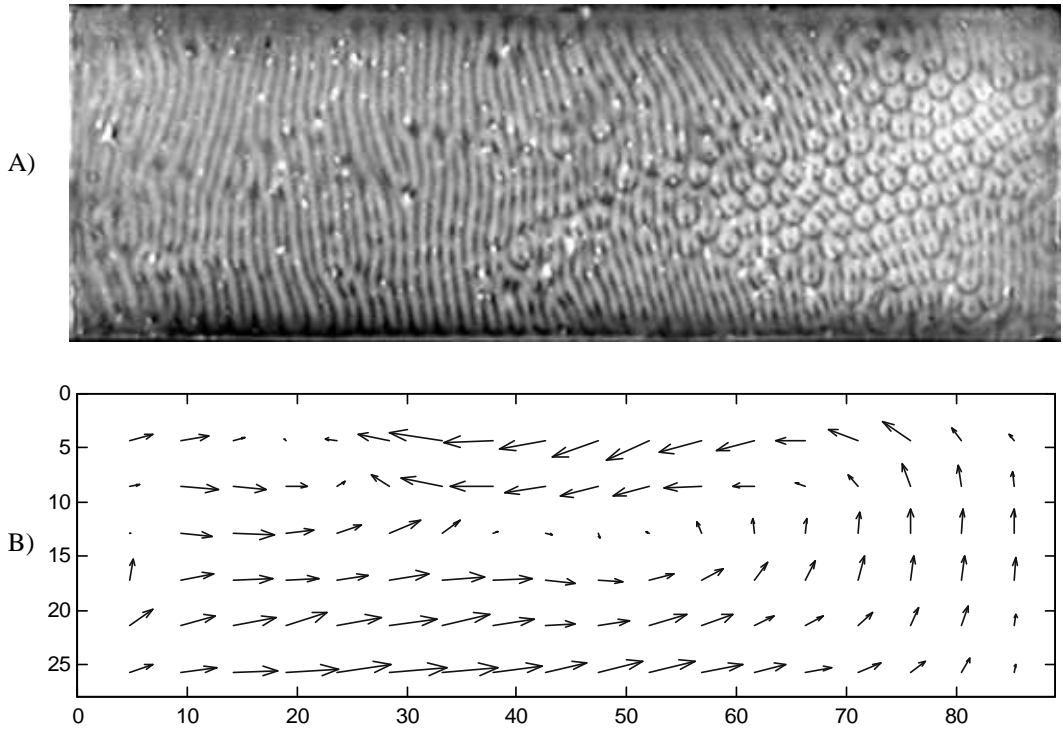

**Fig. 9.** A) Rolls obtained by tilting the cell about an axis parallel to the longer side Ix=1,0º, with H=0, ε=0,7 y d=0,8 mm (case **b**). B) Mean velocities in mm/day for the image A).

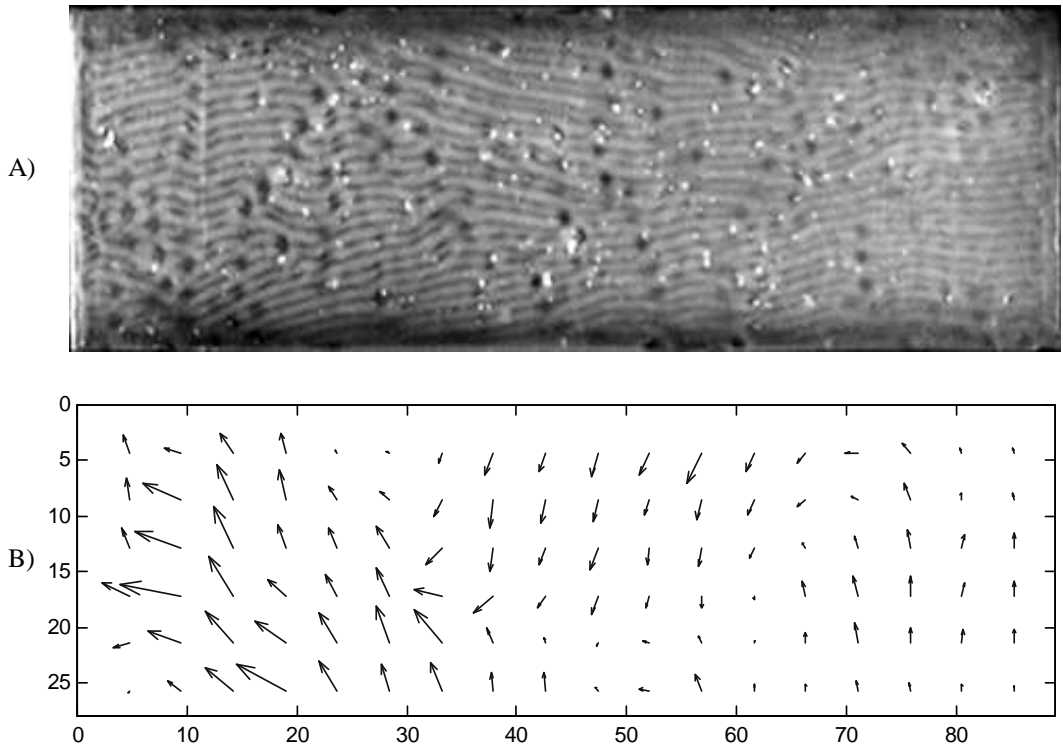

**Fig. 10.** A) Rolls obtained by tilting the cell about an axis parallel to the shorter side Iy=1,0º, with H=0, ε=0,7 y d=0,8 mm (case **d**). B) Mean velocities in mm/day for the image A).

**Table II.** Average and maximum velocities, and their error, for the different cases of inclinations studied.

|    | Thickness (mm) | Angle | $(H/H_c)^2$ | $\varepsilon$ | mean v. (mm/day) | error mean v. | v. max. (mm/day) | error v. max. |
|----|---|---|---|---|---|---|---|---|
| a) | 0,8 | $I_x = 0,5$ | 0,00 | 0,70 | 2,27 | 0,41 | 5,02 | 0,73 |
| b) | 0,8 | $I_x = 1,0$ | 0,00 | 0,70 | 5,23 | 0,71 | 9,97 | 0,58 |
| c) | 0,8 | $I_y = 0,5$ | 0,00 | 0,70 | 1,72 | 0,32 | 5,04 | 0,48 |
| d) | 0,8 | $I_y = 1,0$ | 0,00 | 0,70 | 3,81 | 0,62 | 11,20 | 1,28 |
| e) | 1,0 | $I = 0,0$ | 1,07 | 0,15 | 1,75 | 0,16 | 2,37 | 0,18 |
| f) | 1,0 | $I_u = -0,25$ | 1,07 | 0,15 | 0,40 | 0,07 | 1,15 | 0,12 |
| g) | 1,0 | $I_v = 0,13$ | 1,07 | 0,15 | 0,70 | 0,09 | 1,34 | 0,07 |

Specifically, for angles of about 0.5 degrees, the direction of movement of the structures tended to be perpendicular to the tilt axis (see [28]), from the top of the cell to the lowest side. For inclinations parallel to both sides of the cell, the tilt increased until the destruction of existing structures and their transformation into rolls collinear with the direction of inclination, that is, perpendicular to the tilt axis. This transformation occurred for slopes of 1.0 degree in both cases. In Figure 9A rolls obtained in the case b are observed (Ix = 1), and in Figure 10A that obtained for the case d (Iy = 1). In both cases, the rolls have a wavelength $\lambda_R = 2.08$, slightly higher than that of squares and hexagons. So, when polygonal structures are transformed into rolls, there is a process of expansion of the original structures together with their transformation.

These rolls had a particular dynamic. Moving sideways but not to the same side the entire roll: it was split into pieces, moving a portion of the rolls to the right, and another part to the left. When the rolls were parallel to the shorter side of the cell, that is, in the case b, the movement virtually divided the cell in halves, moving the rolls of the bottom of Figure 9A to the right, and the top to the left. Note that the top of the image coincides with the part that is higher in the cell due to the inclination. For the case d behavior was similar, except that as rolls are positioned parallel to the longer side of the cell, they are fractured into several parts, which move towards one side and the other alternately. Movements are best viewed in Figures 9B and 10B, where the vectors represent the mean velocities of the structures in mm/day, corresponding to the cases b and d respectively.

Since for a horizontal cell of 1.0 mm thick the dynamics was mostly linear with a random initial movement direction (case e), we tried tilting the cell in the way to counteract the movement. To do that, the cell was tilted -0.25° respect to a horizontal axis with components u = (-sin(26.6°), cos(26.6º)), as this is almost perpendicular to the main direction of the overall dynamics. This produced a reduction of the velocities (case f). The next thing we tried was to impose an arbitrary direction to dynamics. For this, the same cell was tilted again 0.125°, this time relative to an axis with components v = (sin(13.3°), cos(13.3°)). The result was that we were able to impose the direction of movement perpendicular to the rotation axis of the structures (case g).

## 4. Discussion and Conclusions

In the field of thermal convective problems in Liquid Crystals, this work presents a new class of transitions of spatial patterns for supercritical convective states. For Homeotropic geometries, previous experiments or theoretical non-linear models do not predict the existence of hexagon-square transitions away from the threshold. So, the discussion of our experiments will be made considering both theory and experimental results, concerning Rayleigh-Benard-Marangoni convection in isotropic newtonian [14] [15] [16] [18] or non-newtonian fluids [29] [30] and instabilities of ferrofluid horizontal films submitted to external perpendicular magnetic fields [27] [31] [26]. In both kinds of systems, hexagon-square transitions are described considering the growth of the second pattern over the first depending on control parameters (temperature differences or intensity of magnetic field).

The first interesting result in our system is the possibility to obtain two kinds of transitions, square-hexagon and hexagon-square, depending on if the critical thermal gradients are smaller or larger. Another interesting result concerns the growth mechanism of square patterns in the hexagon-square transitions. In a similar way as described in some works [18] [27], the mechanism that produce a system with two perpendicular modes with different wave-numbers is the disappearance of one of the three pre-existing mode in hexagonal patterns by displacements of cell rows. In Abou et al. work, several ways have been used to produce hexagon-square transitions and to study the nucleation of the second pattern over the first. The first way was by a slow and continuous increase of magnetic field just to reach the transition. Experiments making a jump in the field intensity revealed the appearance of regular hexagonal patterns when the equilibrium state was reached [27]. Above the onset, they show the coexistence of two considered symmetries, triangular (corresponding to the hexagonal patterns) and square with domains separated by fronts of pentagons. Similarly, in our experiments, this kind of coexistence is observed where the front moves with the same velocity of the structures. As it was remarked before, the displacement of the front is accompanied with the annihilation (or creation) of the vertex in the front generating the squares. The existence of the dynamical effects in our experiments makes it difficult to compare our results with those of ferrofluids. Also, compare our results with previous numerical results is difficult because they consider, in all the cases, only quasi-static hexagon-square transitions with one domain for each structure. However, the qualitatively described propagation of fronts is similar to those observed here.

In this work we present the existence of the traslational pattern movements. Having ruled out the possibility that horizontal thermal gradients were responsible for this dynamics and the possible inclinations with respect to the horizontal position of the cell, it seems more likely that a secondary instability for supercritical thermal gradients exists. However, specific experiments must be performed to answer this question before any conclusions can be drawn.

One of the clearest cases of thermoconvective structures dynamics is observed in binary mixtures, where for the Rayleigh-Bénard problem oscillating rolls are obtained as solutions. If there is also an interface with another fluid and the surface tension becomes play a role in the convection, then hexagons are possible solutions which move in the direction of one of its modes, as in [32]. In that paper three-dimensional equations for Bénard-Marangoni convection in a binary mixture of isotropic fluids are solved by numerical calculation. As the authors point out, this dynamic had not been observed before. That work [32] was published shortly before our previous work [6] [21] where the dynamics that we are studying in depth here was initially announced. Interestingly, the similarity between homeótropos nematic convection and a binary mixture is greater than that described, because in a binary mixture convective instability is found when it is heated above [33], which is not yet registered with any other isotropic fluid.

In the case of inclined cell, one can compare the results with those obtained in the dynamic of thermo-convective rolls observed in inclined cells of isotropic fluid [34]. While our results have similarities with those of Daniels et al., in our study the slope of the cell, not only changed the direction and speed of the structures, but also generated a singular behavior that destroyed hexagons and turned them into rolls, even for angles of inclination extraordinarily small. This is not true for Rayleigh-Bénard in isotropic fluids because the only structure are rolls. In contrast, similar behavior is generated by a horizontal magnetic field. Although both processes are physically different, in both cases there is a horizontal force, gravitational in one case and magnetic in the other, that break symmetry in the xy plane. This symmetry is necessary for the existence of hexagons, and thus both forces generate hexagons - rolls transition. And this transition in our case occurs for angles of only 1 degree, showing how sensitive our system is to perturbations in the inclination.

Here, it was difficult to measure the evolution of wave numbers. Optical methods as spectral techniques are difficult to apply in Nematics because of the strong birefringence of the fluid. However, direct visualisation is easy, although the dynamical effects and non-regularity of the patterns make it difficult, at least in the experiments and geometries considered here, for an accurate quantitative analysis of wave-number evolutions, hysteretic effects, etc..

## 5. Acknowledgements